\documentclass[12pt]{article}
\usepackage{fullpage}
\usepackage{amsmath}
\usepackage{amsthm}
\usepackage{amssymb}
\usepackage{authblk}
\usepackage{tikz}
\usepackage{pgfplots}
\usepackage{url}

\theoremstyle{definition}

\usetikzlibrary{decorations.markings}

\newcommand{\graphnode}[2]{\node[draw,thick,fill=gray,circle,scale=0.4] (#1) at (#2) {}}
\newcommand{\labelednode}[3]{\graphnode{#1}{#2}; \node[below left=-2pt,font=\footnotesize] at (#1) {#3}}
\definecolor{gold}{rgb}{0.85,.66,0}
\tikzset{
midarrow/.append style={decoration={markings, mark=at position #1 with {\arrow{>}}}, postaction={decorate}},
arcweight/.append style={above=-1mm,pos=#1,font=\scriptsize},
}

\begin{document}

\title{A reduced integer programming model for the ferry scheduling problem\footnote{This research work was supported by BC Ferries and MITACS.}}

\author{Daniel Karapetyan}
\author{Abraham P.\ Punnen}
\affil{Department of Mathematics, Simon Fraser University Surrey, Central
City, 250-13450 102nd AV, Surrey, British Columbia, V3T 0A3, Canada\\
{\tt daniel.karapetyan@gmail.com},~~{\tt apunnen@sfu.ca}}
\date{}

\maketitle

\begin{abstract}
We present an integer programming model for the ferry scheduling problem, improving existing models in various ways.  In particular, our model has reduced size in terms of the number of variables and constraints compared to existing models by a factor of approximately $O(n)$, where $n$ being the number of ports.  The model also handles efficiently load/unload time constraints, crew scheduling and passenger transfers.  Experiments using real world data produced high quality solutions in 12 hours using CPLEX~12.4 with a performance guarantee of within 15\% of optimality, on average.  This establishes that using a general purpose integer programming solver is a viable alternative in solving the ferry scheduling problem of moderate size.
\end{abstract}

\section{Introduction}

Given a set of ports, a set of ferries, and a planning horizon, the ferry scheduling problem (FSP) seeks a routing and scheduling scheme for the ferries so that the travel demand emanating at the ports during the planning horizon is satisfied while the operations cost and passenger dissatisfaction are minimized.  Ferry companies are often confronted with the problem of revising the ferry schedules due to a variety of reasons. These include changing demographics within the service area, changes in fleet characteristics, changes in port configuration, altered service level restrictions, responses to customer feedback, and changes in government regulations. Developing an `optimal' schedule is crucial in achieving operational efficiency and  cost savings. Even for a problem that appears to be of a small size (e.g., 4 ferries and 7 ports with a 20 hour planning horizon), the FSP is still complex and requires systematic scientific approaches to construct useful schedules, attain insights into true operational bottlenecks, and to develop proper management strategies.

Optimization problems similar to the FSP has been studied extensively in the operations research literature in the context of airline scheduling~\cite{gt}, public transport scheduling~\cite{c} and scheduling problems in maritime transportation~\cite{Christiansen2004, Christiansen2007}. Such problems can be formulated as integer programming problems. However, solving the resulting integer programming models to optimality  is prohibitively hard most of the time due to the very large size of the  models. Heuristic solution procedures are often employed to obtain practical solutions for such problems. Despite the similarity  with the airline or public transport scheduling problem, special treatment of the FSP is desirable to handle special constraints and to exploit the resulting problem structure.  Further, practical FSPs allow integer programming models of moderate size and this opens new avenues for exploring the usability of powerful general purpose integer programming solvers.

Lai and Lo~\cite{ll} studied the FSP and provided an integer programming formulation with the assumption that all ferries are of same speed.  The paper provides detailed references on related problems and excellent computational results are reported with their heuristic algorithm.  Wang, Lo, and Lai~\cite{wll} considered a slightly more general problem with demand estimation included in the model using appropriate mathematical functions. Mitrovic-Minic and Punnen~\cite{sp} studied the FSP with re-configurable ferries. All these works presented integer programming models and reported that the resulting model was too large to be used in practice.  Thus, they also developed heuristic algorithms.  For other related works on ferry scheduling we refer to~\cite{a3,a2,a1}.  More general reviews of maritime transportation can be found in~\cite{Christiansen2004,Christiansen2007}.

The power of integer programming solvers increased substantially over the past decade. Researchers quantify this improvement by a factor of $7$ approximately~\cite{bixby,hlg,koch,lodi}. Integration of various metaheuristic ideas into modern integer programming solvers not only resulted in powerful general purpose exact solvers, but also improved our ability to use the exact solvers as a heuristic by defining appropriate termination criteria. Moreover, the resulting solution also comes with a proof of quality in terms of percentage deviation from the best lower bound identified. These observations renewed our interest in studying integer programming formulations for the FSP.

In this paper we present an integer programming formulation for the FSP\@.  The size of our formulation (number of variables and constraints) is less than that of the previous models by a factor $O(n)$, where $n$ is the number of ports.  We also rectify some shortcomings in the existing models in terms of crew scheduling, load/unload constraints and passenger transfer constraints.  We have tested our model using real world data, and the outcomes of our experiments indicated that our approach leads to significant savings of computational time, which in turn makes it possible to obtain near-optimal solutions (15\% optimality guarantee for our case study) in practical time.

The paper is organized as follows.  In Section~\ref{sec:existing} we provide a formal definition of the problem and introduce relevant notations.  Section~\ref{sec:mip} deals with our integer programming model.
Summary of experimental results based on our case study is presented in Section~\ref{sec:experiments} followed by concluding remarks in Section~\ref{sec:conclusions}.

\section{Notations and basic model}
\label{sec:existing}

Let us first give a formal definition of the ferry scheduling problem (FSP)\@. Let $P=\{1, 2, \ldots, n \}$ be a set of ports and $F=\{1, 2, \ldots, m \}$ be a set of ferries. For each port $p \in P$, we are given the number of berths $\beta^k$.  For each ferry $f\in F$, we are given its capacity $C^f$, in \emph{automobile equivalent units} (AEQ), and a port $h^f\in P$, called the home port of the ferry $f$. The ferry $f$ starts and ends its service at $h^f$.  We are also given the time $T^f(k, h)$ needed for ferry $f$ to travel from port $k$ to port $h$.\footnote{Our notations follow several simple rules.  The ports are usually denoted by $k$, $s$ or $p$.  Ferries are usually denoted by $f$.  The superscript is used for either port or ferry and subscripts for other indices.}

Let $[\ell, L]$ be the planning horizon. If $\ell = \text{5:00~am}$ and $L = \text{12:00~pm}$, then the planning horizon is the time interval between 5:00~am and 12:00~pm.  A finite travel demand originates at discrete times at a port during the planning horizon with prescribed destination ports.  During the planning horizon, starting from the home port, each ferry visits its destination ports (possibly multiple times)  and returns to its home port.  We call this a \emph{ferry traversal}.  An arrangement of all the arrival and departure times of a ferry $f$ at each of the ports it visits in a traversal is called a \emph{schedule} of the ferry $f$.  A ferry carries passengers as well as vehicles of various types such as trucks, SUVs, buses, cars, other commercial vehicles, etc.  To simplify the demand types, we consider a measure called \emph{automobile equivalent} (AEQ) which is calculated from the itemized demand using a conversion formula.  Thus hereafter, we assume that demand is given in terms of AEQ\@. Each travel demand can be represented by a 4-tuple $(o,d,t,b)$ where $o$ is the origin port, $d$ is the destination port, $t$ is the departure time and $b$ is the demand volume in AEQ\@.  The FSP is to develop schedules for all ferries to move the whole demand volume from the respective origins to destinations so that the weighted sum of total operating cost and total travel time is minimized.

Let us now develop an integer programming formulation for the FSP\@. We follow the standard approach adopted in~\cite{ll,sp,wll} to develop the general structure of the model by making use of ferry flow networks and passenger flow networks.

\subsection{Ferry flow network}
\label{sec:ferry-flow-network}

Our \emph{ferry flow network} is defined as follows.  Let $\delta$ be a given \emph{time increment factor} which is used to discretize the planning horizon.  Let $q$ be the number of time intervals within the planning horizon, i.e., $q \delta = L - \ell$.   Typically, $\ell$ represents 5:00~am, $\delta = 10$~minutes and $q$ is about 120.

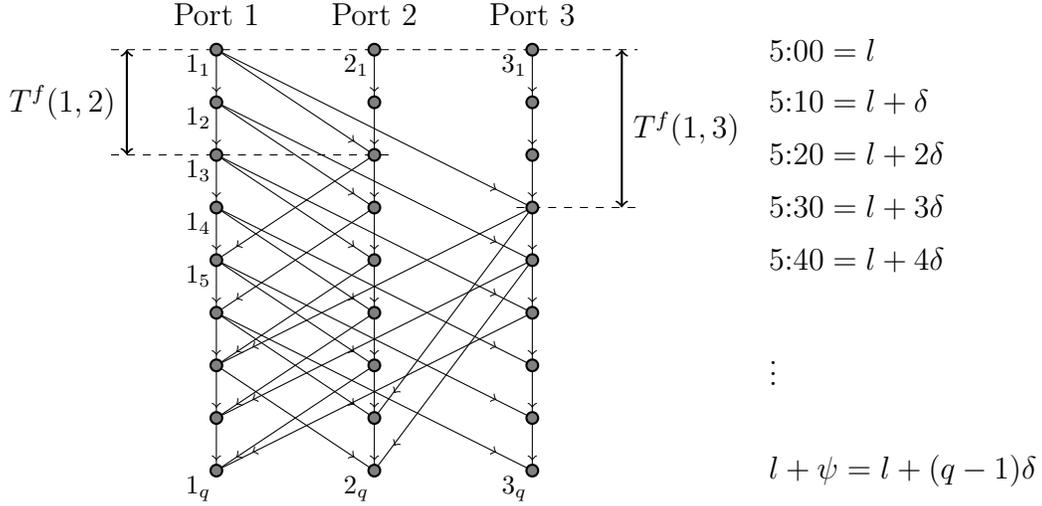
\begin{figure}[ht]
\centering
\begin{tikzpicture}[scale=0.7]
\draw[dashed] (1,8)--(11,8);
\draw[dashed] (1,6)--(6.3,6);
\draw[thick,<->] (1.3,6)--(1.3,8) node[midway,left] {$T^f(1,2)$};

\draw[dashed] (8.7,5)--(11,5);
\draw[thick,<->] (10.7,5)--(10.7,8) node[midway,right] {$T^f(1,3)$};



\foreach \port in {2,3}
{
	\labelednode{v\port-1}{3*\port,8}{$\port_1$};
	\foreach \p in {2,...,8}
		\graphnode{v\port-\p}{3*\port,9-\p};
	\labelednode{v\port-9}{3*\port,0}{$\port_q$};
}

\foreach \p in {1,...,5}
	\labelednode{v1-\p}{3,9-\p}{$1_{\p}$};
\foreach \p in {6,...,8}
	\graphnode{v1-\p}{3,9-\p};
\labelednode{v1-9}{3,0}{$1_q$};


\foreach \i in {1,...,7}
{
	\pgfmathtruncatemacro{\j}{\i + 2};
	\draw[midarrow=0.9] (v1-\i)--(v2-\j);
}

\foreach \i in {1,...,6}
{
	\pgfmathtruncatemacro{\j}{\i + 3};
	\draw[midarrow=0.9] (v1-\i)--(v3-\j);
}
\foreach \i in {3,...,7}
{
	\pgfmathtruncatemacro{\j}{\i + 2};
	\draw[midarrow=0.9] (v2-\i)--(v1-\j);
}
\foreach \i in {4,...,6}
{
	\pgfmathtruncatemacro{\j}{\i + 3};
	\draw[midarrow=0.9] (v3-\i)--(v1-\j);
}
\foreach \i in {4,...,5}
{
	\pgfmathtruncatemacro{\j}{\i + 4};
	\draw[midarrow=0.9] (v3-\i)--(v2-\j);
}

\foreach \port in {1,2,3}
\foreach \i in {1,2,...,8}
{
	\pgfmathtruncatemacro{\j}{\i + 1};
	\draw[midarrow=0.9] (v\port-\i)--(v\port-\j);
}


\foreach \port in {1,2,3}
	\node[above=1ex] at (v\port-1) {Port \port};

\foreach \i/\t in {
	1/5:$00 = l$,
	2/5:$10 = l + \delta$,
	3/5:$20 = l + 2\delta$,
	4/5:$30 = l + 3\delta$,
	5/5:$40 = l + 4\delta$,
	7/\vdots,
	9/$l + \psi= l+(q-1)\delta$}
	\node[right=3cm] at (v3-\i) {\t};
\end{tikzpicture}
\caption{A ferry flow network with Port 1 as home port.  Travel time for the ferry between Port 1 and Port 2 is 20 minutes, Port 1 and Port 3 is 30 minutes, and  port 2 and port 3 is 40 minutes, and $\delta=10$ minutes.  Some service arcs are omitted for clarity of the example.}
\label{fig:basic_network}
\end{figure}

Let $G^f = (V, E^f)$ be the ferry flow network corresponding to ferry $f$.  For each port $k \in P$, let $V^k = \{ k_1, k_2, \ldots, k_q \}$ be the set of nodes in $G^f$ representing `different time-states' of port $k$.  For example, if $\ell = \text{5:00~am}$ and $\delta = 10$ minutes, then $k_1$ represents 5:00~am, $k_2$ represents 5:10~am and so on.  The node set $V$ of $G^f$ is given by $V = \bigcup_{k \in P} V^k$.  The nodes in $V$ can be considered as a rectangular arrangement as depicted in Figure~\ref{fig:basic_network}.  We denote the exact time represented by the node $k_i$ as $\tau(k_i)$.  Thus $\tau(k_i) = \ell + \delta (i - 1)$ for all $k_i \in V$.

Introduce an arc $(k_i,h_j)$ in $E^f$ for $i = 1, 2, \ldots, q$ and $k, h \in P$, $k \neq h$, if a direct service from $k$ to $h$ is allowed and $j \le q$ is the smallest index such that $\tau(h_j) \ge \tau(k_i) + T^f(k, h)$.  Such arcs connecting time-state nodes of two different ports are called \emph{service arcs}.  Also, for each $k \in P$ and $i = 1, 2, \ldots, q - 1$, introduce an arc $(k_i,k_{i+1})$.  These arcs connect two consecutive time state nodes of the same port and are called {\it waiting arcs}.

An example of a ferry flow network $G^f = (V, E^f)$ with three ports is shown in Figure~\ref{fig:basic_network}.  Every service arc $(k_i, h_j) \in E^f$ represents a potential service of ferry $f$ from port $k$ to port $h$ with departure time $\tau(k_i)$.  Every waiting arc $(k_i, k_{i+1})$ represents a waiting of ferry $f$ at port $k$ between times $\tau(k_i)$ and $\tau(k_{i+1})$.

\subsection{Passenger flow network}
\label{sec:passenger-flow-network}

\begin{figure}[ht]
\centering
\begin{tikzpicture}[scale=0.7]

\foreach \port in {1,2,3}
{
	\foreach \p in {1,...,5}
		\graphnode{v\port-\p}{3*\port,6-\p};
	\labelednode{v\port-6}{3*\port,0}{$\port_q$};
	\labelednode{zeta\port}{3*\port,-3}{$\zeta_\port$};
}


\foreach \i in {1,...,4}
{
	\pgfmathtruncatemacro{\j}{\i + 2};
	\draw[midarrow=0.8,dotted] (v1-\i)--(v2-\j);
}

\foreach \i in {1,...,3}
{
	\pgfmathtruncatemacro{\j}{\i + 3};
	\draw[midarrow=0.8,dotted] (v1-\i)--(v3-\j);
}


\foreach \port in {1,2,3}
	\foreach \i in {1,2,...,5}
	{
		\pgfmathtruncatemacro{\j}{\i + 1};
		\draw[midarrow=0.8,dotted] (v\port-\i)--(v\port-\j);
	}

\foreach \port in {3}
	\foreach \i in {1,2,...,6}
		\draw [midarrow=0.3] (v\port-\i) to [out=320,in=40] (zeta\port);

\foreach \k/\h in {1/2,1/3,2/1,2/3,3/1,3/2}
	\draw [midarrow=0.9,thick] (v\k-6) to (zeta\h);


\foreach \port in {1,2,3}
	\node[above=1ex] at (v\port-1) {Port \port};
\end{tikzpicture}
\caption{A passenger flow network $\Omega^3$ with destination port 3.}
\label{fig:network_zeta}
\end{figure}
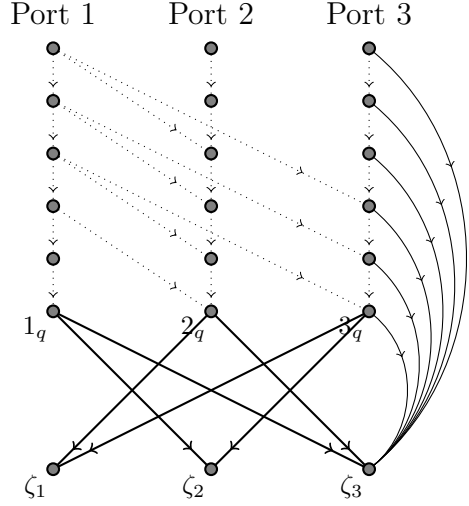

Let $G = (V, E)$ be the minimal supergraph of all $G^f, f \in F$.  Thus, $(i,j) \in E$ implies $(i,j) \in E^f$ for at least one $f$.  It may be noted that $G^f$ will be the same as $G$ for all $f$ if all ferries take the same time to travel between two specified ports.  Introduce $n$ new nodes $\zeta_1,\zeta_2,\ldots ,\zeta_n$ to $G$ and connect node $k_q$ to $\zeta_i$ for all $i\in P\setminus\{k\}$ and for all $k\in P$. The resulting arcs are called \emph{infeasibility arcs}. Also, connect each node $k_i$ for $i=1,2,\ldots ,q$ to $\zeta_k$ by arcs called {\it destination arcs}. Let $\Omega^k = (U,A^k)$ be the resulting graph and we call this the {\it  passenger flow network for destination port $k$}. (See Figure \ref{fig:network_zeta}.) Note that $U=V\cup \{\zeta_1,\zeta_2,\ldots \zeta_n\}$ and $A^k$ consists of $E$ together with all destination arcs for port $k$ and the infeasibility arcs.

Note that we consider all the passengers heading to same port as the same type of flow.  This is different to the traditional approach to maintain a flow for each OD-pair.  Although this enhancement is appropriate in our case study problem, it may not be applicable in some other cases where the information about the origin of a passenger is crucial.

Similar to the ferry flow notation, let $I_k(v) = \{ u :\ (u,v) \in A^k \}$ and $O_k(v) = \{ u :\ (v, u) \in A^k \}$.

\subsection{Basic model}

For any $v \in V$, let $I^f(v) = \{ u :\ (u,v) \in E^f \}$ and $O^f(v) = \{ u :\ (v, u) \in E^f \}$.  For each $(u,v)\in E^f$ consider a 0-1 variable $y_{uv}^f$  defined by
\begin{equation*}
y^f_{uv} = \begin{cases}
1 & \text{if ferry $f$ traverses the arc $(u,v)$,} \\
0 & \text{otherwise.}
\end{cases}
\end{equation*}
Define
\begin{equation}
\label{eq:b}
b^f_v = \begin{cases}
-1 & \text{if $v = h^f_1$,} \\
1 & \text{if $v = h^f_q$,} \\
0 & \text{otherwise.}
\end{cases}
\end{equation}

Let $d^p_{k_i}$ be the travel demand originating at node $k_i$, $k \in P$, $i = 1, 2, \ldots, q$ with destination port $p$. If no travel demand originates at node $k_i$ with destination node $p$, $d^p_{k_i}=0$. The travel demand with destination port $p$ at the node $\zeta_t$ of $\Omega^p$ is given by

\begin{equation*}
d^p_{\zeta_t} = \begin{cases}
\displaystyle{ -\sum_{k\in P}\sum_{i=1}^qd^p_{k_i}} & \text{if $t=p$} \\
0 & \text{otherwise.}
\end{cases}
\end{equation*}

Let $x^k_{uv}$ be the number of passengers (measured in AEQ) traveling along arc $(u,v)$ in $\Omega^k$. Thus the destination port of passengers $x^k_{uv}, (u,v)\in \Omega^k$ is port $k$.

Then we can formulate the FSP as follows:
\begin{align}
\text{Minimize }& \lambda \sum_{f \in F} \sum_{e \in E^f} g^f_e y^f_e  + \nu \sum_{k \in P} \sum_{e \in A^k} c^k_e x^k_e \label{eq:objective}\\
\text{subject to }&\notag\\
&\sum_{v' \in I^f(v)} y_{v'v}^f - \sum_{v' \in O^f(v)} y_{vv'}^f = b^f_v \text{ for every $v \in V$ and $f \in F$,} \label{eq:ferry-flow-balancing}\\
&\sum_{f \in F} y_{k_ik_{i+1}}^f \le \beta^k \text{ for every waiting arc $(k_i, k_{i+1}), k_i \in V^k, k \in P, i \neq q$,} \label{eq:berths}\\
&\sum_{v' \in I_k(v)} x^k_{v',v} - \sum_{v' \in O_k(v)} x^k_{v,v'} = d^k_v \text{ for every $v \in U$ and for every $k \in P$,} \label{eq:passenger-flow-balancing}\\
&\sum_{k\in P} x_{uv}^k \le \sum_{f \in F(u,v)} C^f y_{uv}^f \text{ for every service arc $(u,v)\in \Omega$,} \label{eq:capacity}\\
&\text{$y^f_{uv}\in \{0,1\}$ for $(u,v)\in E^f, f\in F$,}\\
&\text{$x^k_{uv}$ is a non-negative integer for $(u,v)\in A^k, k\in P$,} \label{eq:x}
\end{align}
where $\lambda$ and $\nu$ are coefficients controlling the cost/level of service balance, $g^f_e$ is the cost associated with traversing the arc $e \in E^f$ and $c^k_e$ is the penalty associated with the arc $e \in A^k$.  Let $\mu^f$ be the cost of operating ferry $f$ for one hour when it is in service and $\eta^f$ be the cost of operating ferry $f$ for one hour when it is staying at a port.  These costs include fuel cost and crew salary.  For each edge $e \in E^f$, define $g^f_e$ as
\begin{equation}
\label{eq:g}
g^f_e = (\tau(v) - \tau(u)) \cdot \begin{cases}
\mu^f & \text{if } v, u \in V^k \text{ for some } k \in P,\\
\eta^f & \text{otherwise},
\end{cases}
\end{equation}
where $e = (u, v)$.  Also define $c^k_e$ as
\begin{equation*}
c^k_{uv} = \begin{cases}
M & \text{if $(u,v)$ is an infeasibility arc,} \\
0 & \text{if $(u,v)$ is a destination arc,}\\
\tau(v) - \tau(u)& \text{otherwise,}
\end{cases}
\end{equation*}
where $e = (u, v)$ and $M$ is a very large number. Note that the destination arcs have zero cost and any passenger that does not reach the correct destination port needs to travel through the infeasibility arcs incurring a very large cost.

We require that a ferry needs to be back at its home port at the end of the planning horizon since repositioning cost for ferries is high for our case study whereas for models studied in~\cite{ll,wll} the repositioning cost is assumed to be negligible.

Also note that some of the variables and constraints in the above model are redundant.  Consider, e.g., an arc $(k_1, h_j) \in E^f$, where $k \neq h^f$.  Obviously, this arc can never be traversed and, hence, corresponding variables and constraints can be safely removed.  However, our experiments showed that the preprocessing algorithms embedded into modern integer programming solvers are powerful enough to eliminate such redundancy and, thus, there is no need to complicate the model with such improvements.

\section{Integer programming formulation enhancements}
\label{sec:mip}

The major contribution of this paper is our passenger flow network, see Section~\ref{sec:passenger-flow-network}, that is somewhat different from previous models studied for similar problems.  This allows us to reduce the number of variables and constraints by a factor of $O(n)$ compared to models in~\cite{ll,sp,wll}.  Further, we add additional constraints that allow load/unload time at each port which was not considered in previous models.  This way we can handle berth conflict constraints properly.  Avoiding infeasible passenger transfers is another complicating factor in our model which is also handled by appropriate constraints.  This was not addressed in previous models and for some past studies such an issue was not relevant since passenger transfers were not allowed in those models.  As additional enhancements, we modify the ferry flow network and passenger flow networks to allow more accurate crew scheduling and operating cost calculation. These enhancements to our basic integer programming models are discussed in this section.

\subsection{Loading/unloading times}

The ferry flow balancing constraints (\ref{eq:ferry-flow-balancing}) allow ferry $f$ to pass through the arcs of the ferry flow network and eventually reach the home port at the end of the planning horizon.  However, if ferry $f$ enters a node $k_i$ of $G^f$, $i \neq q$, we need to ensure that it traverses several waiting arcs to allow required loading/unloading time.  The loading/unloading time depends on several factors including ferry layout, port facilities, current load, downloading and uploading traffic, and staff efficiency.  Some of these parameters may vary significantly from time to time and, thus, there exists no accurate way to estimate the time required for a ferry to stay in a port.  Thus, we propose a flexible yet simple way of defining such times that allows fine tuning of the model based on historical data.  We assume that, for each ferry $f$ and each node $k_i$, the minimum loading/unloading time is fixed to $w^f_k \delta$, where $w^f_{k_i}$ is given.  Then, if ferry $f$ traverses a service arc $(h_j, k_i)$, then it also needs to traverse $w^f_{k_i}$ waiting arcs $(k_i, k_{i+1})$, $(k_{i+1}, k_{i+2})$, etc.  This can be ensured by the inequality:
\begin{equation}\label{eq:w1}
|\Delta^f_{k_i}| \sum_{u \in I^f_{\text{serv}}(k_i)}y^f_{u k_i} \leq \sum_{(u,v) \in \Delta^f_{k_i}}y^f_{uv} \text{ for every $k_i \in V$ and $f \in F$} \,,
\end{equation}
where $\Delta^f_{k_i}=\{(k_i,k_{i+1}),(k_{i+1},k_{i+2}),\ldots (k_{r-1},k_{r})\}$, $r=\min\{i+w^f_k,q\}$, and $I_{\text{serv}}^f(k_i) = \{ u :\ (u,k_i) \in E^f, u \notin V^k  \}$.

Given $w^f_{k_i} \delta < T^f(k, h)$ for every $h \in P$, $h \neq k$, the above constraint can be simplified as follows:
\begin{equation}
\label{eq:ferry_waiting_constraint}
\sum_{v' \in O_\text{serv}^f(k_i)} y_{k_i,v'} \le y_{k_j,k_{j+1}} \text{ for every $k_i \in V$ and $f \in F$} \,,
\end{equation}
where $O_\text{serv}^f(v) = \{ u \in O^f(v) :\ (v, u) \text{ is a service arc} \}$ and $j$ is the largest index such that $\tau(k_i) - \tau(k_j) \ge W_k^f$. If no such $j$ exists, the constraint for the corresponding node $k_i$ is not needed.

Some authors add the loading/unloading times to the travel times.  Then a ferry does not need to traverse any waiting arcs and, thus, the berth constraints may be violated.  In our approach, inequalities (\ref{eq:w1}) and (\ref{eq:berths}) handle the load/unload requirements and berth constraints efficiently.

\subsection{Transfer times}

\begin{figure}[ht]
\centering
\begin{tikzpicture}[scale=0.7]
\foreach \port in {1,2,3}
{
	\foreach \p in {1,...,5}
		\graphnode{v\port-\p}{3*\port,4-\p};
}
\draw[midarrow=0.8,thick] (v1-2)--(v2-3);
\draw[midarrow=0.8,thick] (v2-3)--(v2-4);
\draw[midarrow=0.8,thick] (v2-4)--(v1-5);
\draw[midarrow=0.8,thick,dashed] (v3-1)--(v2-2);
\draw[midarrow=0.8,thick,dashed] (v2-2)--(v2-3);
\draw[midarrow=0.8,thick,dashed] (v2-3)--(v3-4);
\end{tikzpicture}
\caption{A sample infeasible transfer from one ferry (solid line) to another ferry (dashed line).}
\label{fig:transfer}
\end{figure}
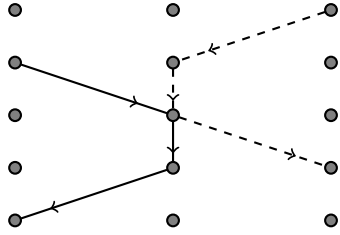

Another important issue that we have to deal with is infeasible transfers. Since there is no penalty for a transfer from one ferry to another, the model favors immediate transfer without allowing time for passengers to unload from one ferry and reload onto another. (See Figure~\ref{fig:transfer} for such an instance.)
Thus, we want to make sure that passengers transferring from one ferry to another one get sufficient unloading and loading time.  Let $T_{k_i} \delta$ be the transfer time required for a passenger arriving at node $k_i$ and $\Delta_{k_i} = \{ k_i, k_{i+1}, \ldots, k_{r-1} \}$, where $r = \min \{ i + T_{k_i}, q \}$.  Let $I_{\text{serv}}(k_i) = \{ u :\ (u,k_i) \in E,\ u \notin V^k \}$.  Then the \emph{transfer constraints} can be stated as
\begin{equation}
\label{eq:transfer_wait}
\sum_{j = i}^t \sum_{u \in I_\text{serv}(k_j)} x^p_{u k_j} \le x^p_{k_t k_{t+1}} \text{ for $t = i$ to $r - 1$, $k_i \in V, i\neq q$ and $p \in P \setminus \{ k \}$.}
\end{equation}

Note that (\ref{eq:transfer_wait}) affects not only the transferring passengers but also the passengers remaining on a ferry.  Thus, we require $T_{k_i} \le \min_{f \in F} w^f_{k_i}$ for each $k_i \in V$.

The transfer constraints force that the flow with destination port $p$ along each of the arcs $(k_j,k_j+1)$, $k_j \in \Delta_{k_i}$, is at least the total flow with destination port $p$ arriving at node $k_i$ through service arcs. Note that we have $|\Delta_{k_i}|$ transfer inequalities for each $k_i\in V$, $k\in P$. If port $k$ has only one berth or $T_{k_i}=1$, we can replace the $|\Delta_{k_i}|$ transfer inequalities at node $k_i$ by a single inequality
\begin{equation}
\label{eq:transfer1}
\sum_{u \in I_\text{serv}(k_i)} x^p_{u k_i} \le x^p_{k_ik_{i+1}} \text{ for every $k_i \in V,\ i \neq q$ and $p \in P$.}
\end{equation}

\subsection{Updated model}

The foregoing discussions can be summarized in a formal integer programming formulation (\ref{eq:objective})--(\ref{eq:x}) extended by including the constraint sets (\ref{eq:w1}) and (\ref{eq:transfer_wait}).


The resulting model has $O(q m n^2)$ constraints and $O(qmn^3)$ variables.  If we used the demand type as OD-pairs, as considered in~\cite{ll,wll,sp}, instead of passengers with same destination port considered in the present model, the number of constraints and variables would have been $O(q m n^3)$ and $O(qmn^4)$ respectively, i.e., the model size in our case is smaller by a factor of $O(n)$.  As it was mentioned above, the drawback of such an approach is that the model becomes less flexible.  Indeed, now we cannot distinguish passengers traveling to the same destination in terms of their origin ports.  However, in our case it is not necessary because the model allows us to calculate the total traveling time of all the passenger. The major advantage, however, is the reduced problem size which makes the usage of general purpose integer programming solvers a viable alternative.

Note that in some studies (see, e.g., \cite{a2}), the problem size depends not on the number of ferries but on the number of ferry types.  In our case study, all the vessels were of different types and, thus, we decided not to complicate the discussion by allowing several ferries of the same type.  However, such a modification can be easily incorporated in the proposed model.

\subsection{Operational cost refinements and crew scheduling}
\label{sec:crews}





Recall that for a waiting arc $e = (k_i, k_{i+1})$, the operational cost is defined as $g^f_e = \delta \mu^f$, see (\ref{eq:g}). If all the ferries start operation at the beginning of the planning horizon and terminate at the end of the planning horizon, this is an appropriate cost assignment.  However, this need not be true in some applications. For example, a ferry may start at 10:00~am and terminate at 4:00~pm. If we use the model from the previous section directly, we are counting the excess cost of the waiting arcs from 5:00~am to 10:00~am and from 4:00~pm to 11:00~pm which may alter the optimal solution.  However, the correct calculation of costs for the waiting arcs in such cases is possible by a minor change in our model.

\begin{figure}[h!]
\centering
\begin{tikzpicture}[scale=0.7]


\labelednode{alpha}{-3,6}{$\alpha^f$};
\labelednode{beta}{-3,1}{$\beta^f$};

\foreach \port in {1,2,3}
	\foreach \p in {1,...,8}
		\graphnode{v\port-\p}{3*\port,8-\p};


\foreach \i in {1,...,6}
{
	\pgfmathtruncatemacro{\j}{\i + 2};
	\draw[midarrow=0.8,dotted] (v1-\i)--(v2-\j);
}

\foreach \i in {1,...,5}
{
	\pgfmathtruncatemacro{\j}{\i + 3};
	\draw[midarrow=0.8,dotted] (v1-\i)--(v3-\j);
}


\foreach \port in {1,2,3}
\foreach \i in {1,2,...,7}
{
	\pgfmathtruncatemacro{\j}{\i + 1};
	\draw[midarrow=0.8,dotted] (v\port-\i)--(v\port-\j);
}

\foreach \p in {1,...,7}
	\draw[midarrow=0.8] (alpha)--(v1-\p);

\foreach \p in {2,...,8}
	\draw[midarrow=0.7] (v1-\p)--(beta);


\foreach \port in {1,2,3}
	\node[above=1ex] at (v\port-1) {Port \port};
\end{tikzpicture}
\caption{Ferry flow network with nodes $\alpha^f$ and $\beta^f$ that allow better calculation of the crew salaries and loading/unloading stays.  In this example, Port 1 is the home port $h^f$ of ferry $f$.}
\label{fig:network_alpha_beta}
\end{figure}
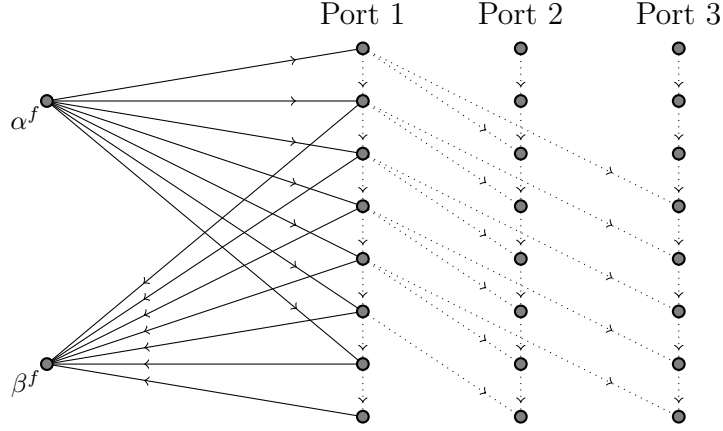

In the ferry flow network $G^f$, $f \in F$, introduce two new nodes $\alpha^f$ and $\beta^f$ and connect $\alpha^f$ to all nodes in $V^{h^f} \setminus \{ h^f_q \}$.  The resulting arcs $(\alpha^f,h^f_i)$, $i=1,2,\ldots,q-1$, are called \emph{in-port arcs}. Likewise, connect each node of $V^{h^f} \setminus \{ h^f_1 \}$ to $\beta^f$ and the resulting arcs $(h^f_i,\beta^f)$, $i=2,3,\ldots, q$, are called {\it out-port arcs}. (See Figure \ref{fig:network_alpha_beta}.)

Set $g^f_e = 0$ for each in-port and out-port arc.  Let ferry $f$ start from node $\alpha$ and terminate at node $\beta$.  Consequently we change the definition of $b^f_v$ (see (\ref{eq:b})) for this modified $G^f$ as
\begin{equation*}
b^f_v = \begin{cases}
-1 & \text{if $v = \alpha^f$,} \\
1 & \text{if $v = \beta^f$,} \\
0 & \text{otherwise.}
\end{cases}
\end{equation*}
Now the operational cost of a ferry $f$ excludes the time spent at the home port before and after its operation.

\bigskip

Let us now further develop the model to handle a situation when the crew time per planning horizon is divided in two shifts, a morning shift and an evening shift. Let $\sigma^f$ be the total crew salary for ferry $f$ per shift. We are given a time window $[t_1,t_2]$, and during this period the ferries must reach their home port for a crew exchange. One way to deal with this problem is to divide the planning horizon into two; $[\ell, t_1]$ and $[t_2,L]$. However, this approach creates another issue where we force passengers to reach their destination during the planning horizon $[\ell,t_1]$ and move demands that could not reach the destination to the planning horizon $[t_2,L]$.  This is likely to make the resulting solution suboptimal. We now show how we can make simple modifications in the definition of $G^f$ to achieve the desired outcome without breaking the problem into two and compromising for suboptimal solutions.

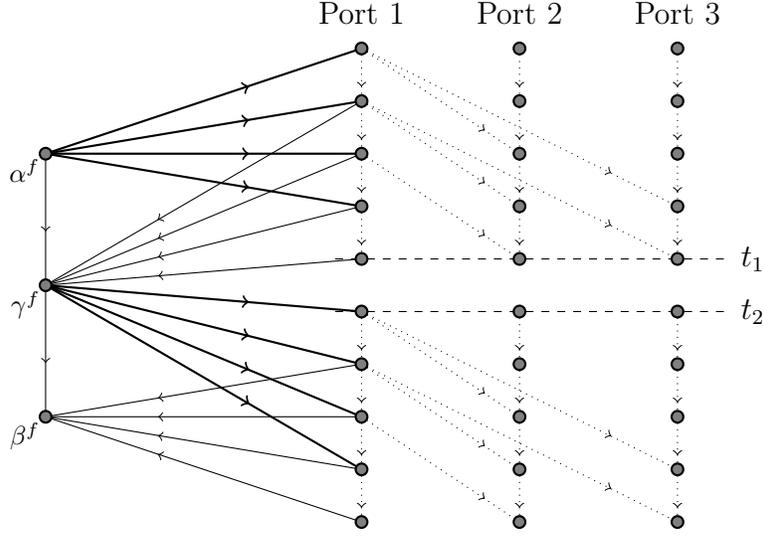
\begin{figure}[ht]
\centering
\begin{tikzpicture}[scale=0.7]

\labelednode{alpha}{-3,7}{$\alpha^f$};
\labelednode{gamma}{-3,4.5}{$\gamma^f$};
\labelednode{beta}{-3,2}{$\beta^f$};

\draw[thin,dashed] (2.5,4)--(10,4) node[right]{$t_2$};
\draw[thin,dashed] (2.5,5)--(10,5) node[right]{$t_1$};

\foreach \port in {1,2,3}
	\foreach \p in {1,...,10}
		\graphnode{v\port-\p}{3*\port,10-\p};


\foreach \i in {1,2,3,6,7,8}
{
	\pgfmathtruncatemacro{\j}{\i + 2};
	\draw[midarrow=0.8,dotted] (v1-\i)--(v2-\j);
}

\foreach \i in {1,2,6,7}
{
	\pgfmathtruncatemacro{\j}{\i + 3};
	\draw[midarrow=0.8,dotted] (v1-\i)--(v3-\j);
}


\foreach \port in {1,2,3}
	\foreach \i in {1,...,4,6,7,...,9}
	{
		\pgfmathtruncatemacro{\j}{\i + 1};
		\draw[midarrow=0.8,dotted] (v\port-\i)--(v\port-\j);
	}

\draw[midarrow=0.6] (alpha)--(gamma);
\draw[midarrow=0.6] (gamma)--(beta);

\foreach \p in {1,...,4}
	\draw[thick, midarrow=0.65] (alpha)--(v1-\p);

\foreach \p in {7,...,10}
	\draw[midarrow=0.65] (v1-\p)--(beta);

\foreach \p in {2,...,5}
	\draw[midarrow=0.65] (v1-\p)--(gamma);

\foreach \p in {6,...,9}
	\draw[thick, midarrow=0.65] (gamma)--(v1-\p);


\foreach \port in {1,2,3}
	\node[above=1ex] at (v\port-1) {Port \port};

\end{tikzpicture}
\caption{Ferry flow network with nodes $\alpha^f$, $\beta^f$ and $\gamma^f$ that allow better calculation of the crew salaries and correct calculation of the fuel charges.  In this example, Port 1 is the home port $h^f$ of ferry $f$.}
\label{fig:network_alpha_beta_gamma}
\end{figure}

In the original ferry flow network $G^f$ as introduced in Section~\ref{sec:ferry-flow-network}, add three nodes $\alpha^f, \beta^f$ and $\gamma^f$. Connect $\alpha^f$ to $h^f_i\in V^{h^f}$ by an arc whenever $\tau(h^f_i) < t_1$ and set the cost of this arc to $\sigma^f$. Similarly, connect $h^f_i$ to $\gamma^f$ by an arc $(h^f_i, \gamma^f)$ of zero weight for every $i > 1$ such that $\tau(h^f_i) \le t_1$. Set the cost $g^f_e$ of the waiting arcs $e$ involving time state nodes $v$ with $t_1 \leq \tau(v) \leq t_2$ to a very large number $M$\footnote{We assume that such arcs exist.}. Now connect $\gamma^f$ to $h^f_i$ by an arc $(\gamma^f, h^f_i)$ of weight $\sigma^f$ for every $i < q$ such that $\tau(h^f_i) \ge t_2$. Connect each node $h^f_i$ to $\beta^f$ by an arc $(h^f_i, \beta^f)$ of zero weight for every $i$ such that $\tau(h^f_i) > t_2$.  Finally connect $\alpha^f$ to $\gamma^f$ and $\gamma^f$ to $\beta^f$ by arcs of zero weights, see Figure~\ref{fig:network_alpha_beta_gamma}.

The modified definition of $G^f$ can easily be incorporated into our integer programming model.  It can be verified that the modification discussed here correctly handle the desired outcome.  In fact, this is the model we used in our case study experiments.

\section{Experimental analysis}
\label{sec:experiments}

Our integer programming model, as modified in the previous section, was tested using data provided by a major ferry company.  The company operated 3 to 4 ferries linking 7 ports located at various islands.  For peak season, 4 ferries were used but for off-peak season only 3 ferries were used. The demand data was obtained from ticket selling statistics.  Each ticket is sold at the berth for a particular service and, thus, the data was accurate.  We only consider vehicles to define the demand value.  However, if no vehicles were transferred between two ports but there were some foot passengers, we assume the demand volume to be 1~AEQ to ensure that these passengers will be taken into account by the model.  The total demand in the considered problems was 1000 to 2000~AEQ per day.

To solve the integer programming model, we used CPLEX~12.4 where the problem input was prepared using a C\# code. The experiments were run on a Windows 7 PC with an Intel i7 3.4~GHz CPU (8 virtual cores) and 16~GB of memory.  The values of various parameters for the model were set as follows: $\delta = 10$~minutes, $\ell = \text{5:00 am}$ and $L = \text{12:00 midnight}$.  The values for shift cost $\sigma^f$ for ferry $f$ and operating costs for these ferries were provided by the ferry company.

In order to speed up calculations, we provided a trivial starting solution for each problem.  Such a solution defines a schedule, where each ferry stays at the home port for the whole day.  Due to the infeasibility arcs, such a solution is feasible although none of the passengers reach their destination.  When possible, we also tried to use the existing schedules as the starting solutions.  The solver in such cases was not able to improve the schedule except some very minor changes and spent almost the same time as if no solution is provided to prove the quality of the schedule.

The travel demand varies with respect seasons and days of the week.  To validate our model, we selected Off-Peak Monday--Thursday and Peak Friday demands.  The former is interesting because it is operated around 150 days per year.  The latter is one of the busiest days and, thus, requires maximum capacity.  Since the demand data is taken from ticket selling statistics, the traffic volume is essentially what is carried by the current schedule and passengers arrived at the ports as per the prescribed schedule.  Thus, in terms of the proposed model, it is quite likely that the current schedule is close to optimum, if not optimum.  Hence, by comparing the original schedules with the schedules generated by our model for the original fleet configuration, we are able to evaluate our method.

\begin{figure}[h]\begin{tikzpicture}
\tikzset{
every minor grid/.style={dashed,thin,lightgray},
every major grid/.style={solid,thin,lightgray}
}
\begin{axis}[
clip=false,
legend style={cells={anchor=west}, anchor=south, at={(0.5,1.06)}},
legend columns = 4,
width=\textwidth,
height=\textheight,
tick align=outside,
yminorgrids=true,
ymajorgrids=true,
xmajorgrids=true,
xminorticks=false,
xmajorticks=true,
y dir=reverse,
axis x line*=bottom,
axis y line*=left,
xmin=0.5,xmax=7.5,ymin=240,ymax=1500,
minor y tick num=5,
xtick={1,2,3,4,5,6,7},
xticklabels={A,B,C,D,E,F,G},
ytick={0,60,120,180,240,300,360,420,480,540,600,660,720,780,840,900,960,1020,1080,1140,1200,1260,1320,1380,1440},
yticklabels={00:00,01:00,02:00,03:00,04:00,05:00,06:00,07:00,08:00,09:00,10:00,11:00,12:00,13:00,14:00,15:00,16:00,17:00,18:00,19:00,20:00,21:00,22:00,23:00,24:00},
]
\addplot[{mark=*,mark size=1.5pt,mark options={opacity=0.7}}] coordinates {
(0.93, 320)
(2.93, 370)
(2.93, 380)
(4.93, 410)
(4.93, 420)
(1.93, 470)
(1.93, 480)
(0.93, 520)
(0.93, 550)
(5.93, 620)
(5.93, 630)
(2.93, 670)
(2.93, 680)
(1.93, 710)
(1.93, 720)
(0.93, 760)
(0.93, 860)
(1.93, 900)
(1.93, 920)
(0.93, 960)
(0.93, 990)
(1.93, 1030)
(1.93, 1050)
(2.93, 1080)
(2.93, 1120)
(0.93, 1170)
(0.93, 1200)
(1.93, 1240)
(1.93, 1250)
(0.93, 1290)
};
\addlegendentry{Ferry 127}
\addplot[{very thick,dotted,mark=square,mark size=1.5pt,mark options={solid}}] coordinates {
(0.976666666666667, 310)
(5.97666666666667, 380)
(5.97666666666667, 390)
(2.97666666666667, 430)
(2.97666666666667, 450)
(0.976666666666667, 500)
(0.976666666666667, 520)
(1.97666666666667, 560)
(1.97666666666667, 570)
(0.976666666666667, 610)
(0.976666666666667, 640)
(4.97666666666667, 710)
(4.97666666666667, 720)
(2.97666666666667, 750)
(2.97666666666667, 760)
(0.976666666666667, 810)
(0.976666666666667, 850)
(5.97666666666667, 920)
(5.97666666666667, 930)
(4.97666666666667, 990)
(4.97666666666667, 1000)
(0.976666666666667, 1070)
(0.976666666666667, 1100)
(2.97666666666667, 1150)
(2.97666666666667, 1160)
(4.97666666666667, 1190)
(4.97666666666667, 1200)
(5.97666666666667, 1260)
(5.97666666666667, 1270)
(0.976666666666667, 1340)
};
\addlegendentry{Ferry 70 (1)}
\addplot[{dashed,mark=diamond*,mark size=2pt,mark options={solid,draw=black}}] coordinates {
(7.02333333333333, 850)
(4.02333333333333, 950)
(4.02333333333333, 960)
(5.02333333333333, 1010)
(5.02333333333333, 1020)
(7.02333333333333, 1080)
(7.02333333333333, 1110)
(5.02333333333333, 1170)
(5.02333333333333, 1180)
(4.02333333333333, 1230)
(4.02333333333333, 1240)
(7.02333333333333, 1340)
};
\addlegendentry{Ferry 70 (2)}
\addplot[{very thick,mark=triangle*,mark size=2pt,mark options={opacity=0.7}}] coordinates {
(4.07, 400)
(2.07, 440)
(2.07, 450)
(3.07, 480)
(3.07, 490)
(5.07, 520)
(5.07, 530)
(7.07, 590)
(7.07, 620)
(5.07, 680)
(5.07, 700)
(3.07, 730)
(3.07, 740)
(2.07, 770)
(2.07, 780)
(4.07, 820)
(4.07, 860)
(7.07, 940)
(7.07, 990)
(5.07, 1050)
(5.07, 1060)
(3.07, 1090)
(3.07, 1110)
(7.07, 1180)
(7.07, 1230)
(3.07, 1300)
(3.07, 1310)
(2.07, 1340)
(2.07, 1350)
(4.07, 1390)
};
\addlegendentry{Ferry 192}
\end{axis}
\begin{axis}[
axis x line*=top,
axis y line*=right,
width=\textwidth,
height=\textheight,
tick align=outside,
y dir=reverse,
xmin=0.5,xmax=7.5,ymin=240,ymax=1500,
minor y tick num=5,
xtick={1,2,3,4,5,6,7},
xticklabels={A,B,C,D,E,F,G},
ytick={0,60,120,180,240,300,360,420,480,540,600,660,720,780,840,900,960,1020,1080,1140,1200,1260,1320,1380,1440},
yticklabels={00:00,01:00,02:00,03:00,04:00,05:00,06:00,07:00,08:00,09:00,10:00,11:00,12:00,13:00,14:00,15:00,16:00,17:00,18:00,19:00,20:00,21:00,22:00,23:00,24:00},
]
\end{axis}
\end{tikzpicture}
\caption{A sample schedule produced by our model.}
\label{fig:schedule_friday1}
\end{figure}
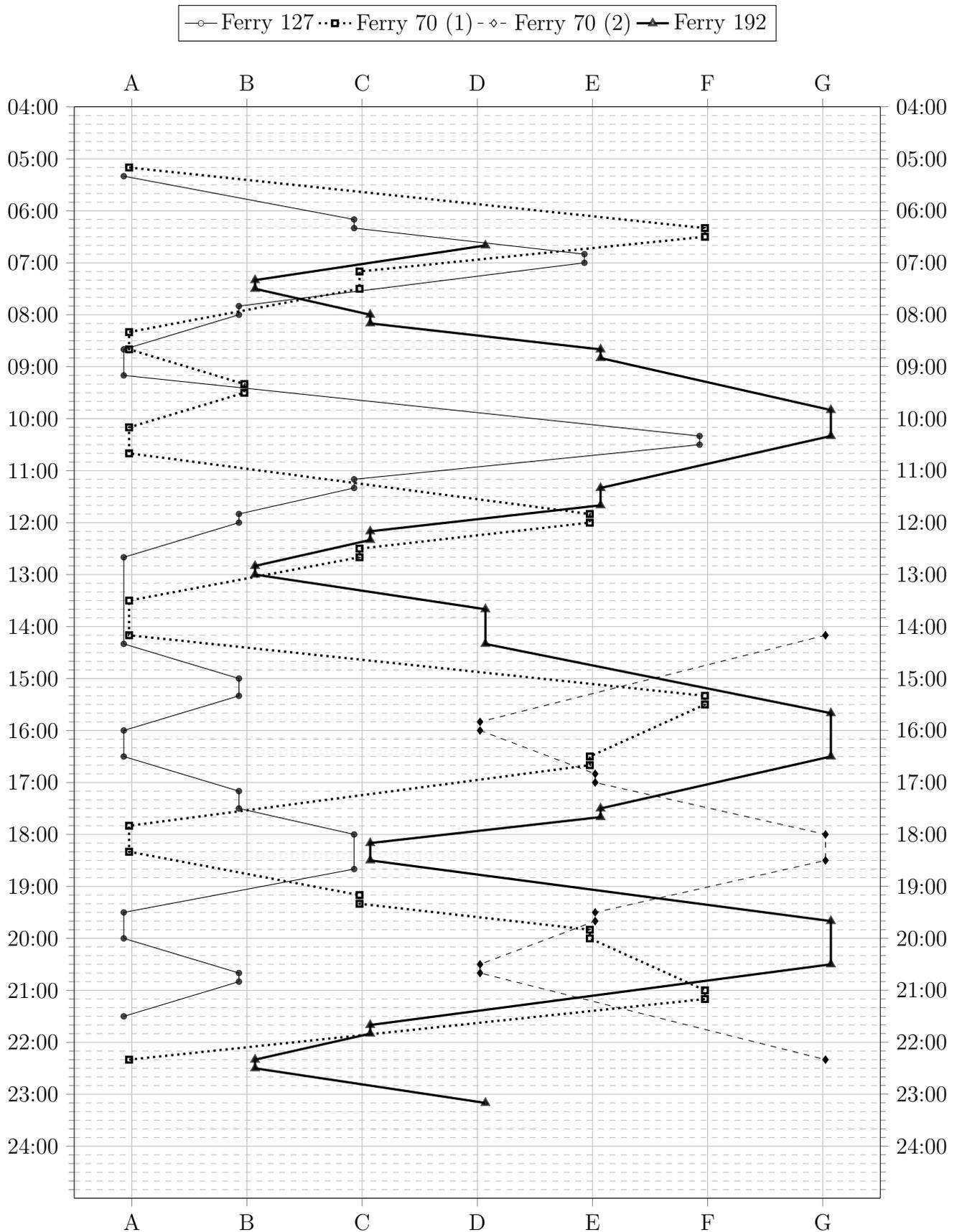

We allowed 12 hours to solve each problem, and the model produced schedules almost identical to the current schedule of the case study and provided a performance guarantee of within 15\% of optimality (in terms of the objective function value), on average.  For example, for the current fleet configuration and the Peak Friday demand data, the routes produced by our model (see Figure~\ref{fig:schedule_friday1}) have only two minor changes comparing to the original routes, and these changes are arguably beneficial.  The departure and arrival times in the produced schedule are also very close to the original schedule; the difference is mainly caused by the lower time resolution of our model (10 minutes) comparing to the currently implemented schedules (5 minutes).  The model contained around $4.7 \cdot 10^4$ integer variables and $4.2 \cdot 10^4$ constraints but its size was reduced by CPLEX preprocessor to $3.8 \cdot 10^4$ integer variables and $3.3 \cdot 10^4$ constraints.  Observe that the real number of variables in the model is well below the estimation $O(qmn^3)$.  That is because the speeds of the ferries considered in our case study are very similar, and many of the arcs in the ferry flow networks coincide, which in turn reduces the size of the passenger flow networks.

For the other fleet configurations, the model was able to produce schedules of reasonable operational costs and service levels.  Our computational experiments validate the model and establish that general purpose integer programming solvers could be a viable alternative to solve moderate sized ferry scheduling problem.

\section{Conclusions}
\label{sec:conclusions}

In this paper, we developed an integer programming model for the ferry scheduling problem. Our model has reduced size compared to existing models by a factor of approximately $O(n)$. The model handles efficiently such new constraints as crew scheduling, loading/unloading time and passenger transfer. The model was able to produce high quality solutions in 12 hours using CPLEX~12.4.

\newpage\pagestyle{empty}


\begin{thebibliography}{9}
\bibitem{bixby}R.E. Bixby. Mixed integer programming: it works better than you may think. Gurobi optimization.  
\bibitem{c} A. Ceder. Designing public transport network and routes.
In \emph{Advanced Modeling for Transit Operations and Service Planning}, W.~Lam and M.~Bell (Eds.),
Elsevier, 2003.
\bibitem{Christiansen2004} M. Christiansen, K. Fagerholt, and D. Ronen. Ship Routing and Scheduling: Status and Perspectives. \emph{Transportation Science}, 38(1), 1--18, 2004.
\bibitem{Christiansen2007} M. Christiansen, K. Fagerholt, B. Nygreen, and D. Ronen. Maritime Transportation. In \emph{Handbook in Operations Research and Management Science}, C.~Barnhart and G.~Laporte (Eds.), vol.~14, pp. 189--284, Elsevier, 2007.
\bibitem{gt}R. Gopalan and K.T. Talluri. Mathematical models in airline schedule planning: A survey. \emph{Annals of Operations Research} 76 (1998) 155-185.
\bibitem{hlg}L.M. Hvattum, A. Lokketangen, and F. Glover. Comparisons of commercial MIP solvers and an adaptive memory (tabu search) procedure for a class of 0-1 integer programming problems. \emph{ Algorithmic Operations Research} 7 (2012) 13--20.
\bibitem{a3}W.M.K.A.W. Zaimi, M.S. Abu, A.K. Junoh, W.N.M. Ariffin. Ferry Scheduling Model Using Linear Programming Technique. In \emph{proc.\ of the 3rd International Conference on Computer Research and Development (ICCRD)}, 4 (2011) 337--341.
\bibitem{koch}T. Koch, T. Achterberg, E. Anderson, O. Bastert, T. Berthold, R.E. Bixby, E. Danna, G. Gamrath, A.M. Gleixner, S. Heinz, A. Lodi, H. Mittelmann, T. Ralphs, D. Salvagnin, D.E. Steffy, and K. Wolter.  MILP 2010 --- Mixed integer programming library version 5. \emph{Mathematical Programming Computation} 3 (2011) 103--163.
\bibitem{ll} M.F. Lai and H.K. Lo.  Ferry service network design: optimal fleet size, routing, and scheduling. \emph{Transportation research PART A: Policy and Practice} 38 (2004) 305--328.
\bibitem{ll} J.T. Linderoth and A. Lodi.  MILP Software.  In \emph{Wiley Encyclopedia of Operations Research and Management Science}, J.J. Cochrane (Ed.).
\bibitem{lodi} A. Lodi.  MIP computation and beyond.  Technical report ARRIVAL-TR-0229.
\bibitem{sp} S. Mitrovic-Minic and A.P. Punnen.  Routing and scheduling of a heterogeneous fleet of re-configurable ferries: a model, a heuristic, and a case study. International Conference on Operations Research, Zurich, Switzerland, 2011.
\bibitem{wll}  Z.W. Wang, H. Lo, and M.F. Lai.  Mixed-fleet Ferry Routing and Scheduling.  \emph{Lecture Notes in Economics and Mathematical Systems 600: Computer-Aided Systems in Public Transport}, M.~Hickman, P.~Mirchandani, and S.~Voss (Eds.), 181--194, 2008.
\bibitem{a2}D.Z.W. Wang and H.K. Lo.  Multi-fleet ferry service network design with
passenger preferences for differential services. \emph{Transportation Research Part B} 42 (2008) 798--822.
\bibitem{a1} S. Yan, C.-H. Chen, H.-Y. Chen, and T.-C. Lou.  Optimal scheduling models for ferry
companies under alliances.  \emph{Journal of Marine Science and Technology} 15 (2007) 53--66.

\end{thebibliography}
\end{document}